\documentclass[prl,aps,showpacs,twocolumn]{revtex4}
\usepackage{graphicx}

\begin{document}

\title{Light bullets in quadratic media with normal dispersion at the second
harmonic}
\author{I. N. Towers}
\altaffiliation[Current address: ]{School of Mathematics
and Statistics, University of New South Wales at ADFA 2600, ACT, Australia}
\author{B. A. Malomed}
\affiliation{Department of Interdisciplinary Studies, Faculty of
Engineering, Tel Aviv University, Tel Aviv 69978, Israel}
\author{F. W. Wise}
\affiliation{Department of Applied Physics, 212 Clark Hall,
Cornell University, Ithaca, NY 14853}

\begin{abstract}
Stable two- and three-dimensional spatiotemporal solitons (STSs)
in second-harmonic-generating media are found in the case of
\textit{normal} dispersion at the second harmonic (SH). This
result, surprising from the theoretical viewpoint, opens a way for
experimental realization of STSs. An analytical estimate for the
existence of STSs is derived, and full results, including a
complete stability diagram, are obtained in a numerical form. STSs
withstand not only the normal SH dispersion, but also finite
walk-off between the harmonics, and readily self-trap from a
Gaussian pulse launched at the fundamental frequency.
\end{abstract}
 \maketitle

A multidimensional soliton is a self-supporting solitary pulse
resulting from the balance between nonlinearity of the medium and
diffraction and dispersion of the wave field. The most physically
relevant realization of this is provided by optical spatiotemporal
solitons (STSs), alias ``light bullets'', which are self-confined
in the longitudinal and transverse directions \cite{Yaron}.

Spatial optical solitons, which are not localized in the
longitudinal direction, have been studied in depth \cite{review},
and observed in both one- \cite{1d} and two- \cite{2d} dimensional
(1D and 2D) geometries. Quasi-2D STSs, which are localized in the
longitudinal and one transverse directions, have been observed too
\cite{Liu99}. A challenge for the experimentalist is creating a
fully 3D bullet. The issue has a special physical purport, as,
unlike 2D solitons which occur in various media, optical crystals
provide for a \emph{single} real possibility to create 3D solitons
in classical physics. Besides their significance as fundamental
objects, STSs may provide THz switching rates in optic-logic
systems, if used as information bits \cite{Macleod}.

It is well known that STSs are unstable in Kerr media
\cite{Rasmussen}. They can be stabilized if the nonlinearity is
saturable \cite{Blagoeva}. Other effects can also stabilize a
spatiotemporal pulse, as shown in an experiment demonstrating
stable space-time focusing in a planar waveguide \cite{Eisenberg}.
However, it was not a soliton, as the stabilization was due to
multi-photon absorption and intra-pulse Raman scattering, which
are dissipative effects.

A promising path for making stable STSs is to employ
second-harmonic generation (SHG). Various types of solitons in SHG
materials have attracted a lot of interest \cite{Buryak}. In
particular, it was proven long ago \cite {Kanashov}, by means of
variational estimates, that SHG must give rise to \emph{stable} 2D
and 3D solitons realizing a minimum of the Hamiltonian.

Most theoretical works on STSs in SHG media assumed spatiotemporal
isotropy \cite{Skr}, which implies equal GVD
(group-velocity-dispersion) coefficients at the fundamental
frequency (FF) and second harmonic (SH), that does not take place
in reality. A general non-symmetric case was considered in Ref.
\cite{Malomed97}. Using the variational approximation (VA), STSs
were predicted, and numerical simulations verified their stable
existence in the 2D case. Later, numerically exact 2D and 3D
soliton solutions were found and their stability was confirmed
\cite{Mihalache98}. Then, another ingredient important for the
description of the real physical situation was introduced, namely,
the group-velocity mismatch (GVM, alias walkoff) between the
harmonics, showing that stable STSs exist in this case too \cite
{Mihalache99a}.

It is obvious that solitons in SHG media may only exist if GVD is
anomalous at FF. Even without postulating the above-mentioned
spatiotemporal isotropy, \emph{all} the previous works assumed
that GVD must be anomalous at SH too. However, this assumption
ignores a fundamental problem: in available SHG materials, FF
suffers strong absorption if the SH wavelength is long enough for
the dispersion to be anomalous. Note that, in Ref.
\cite{Malomed97}, VA formally predicted STSs for normal GVD at SH,
but simulations led to conclusion that it could really exist only
with anomalous GVD at SH, while the prediction for the normal SH
dispersion was an artifact of VA based on the Gaussian ansatz,
which incorrectly treats exponentially decaying tails of the
soliton (see a discussion of the role of the tails below). In the
3D case, VA predicted that STSs could exist at small normal values
of GVD at SH, but this was never checked. Besides its crucial
importance for the creation of LBs in experiment, the issue is
also a challenge for analysis of multidimensional nonlinear-wave
models.

In this work we demonstrate that, \emph{contrary} to the common
belief, STSs in the 3D and 2D cases effectively exists and can be
stable in SHG media with normal GVD at SH. We also conclude that
it survives in the presence of GVM. These results greatly enhance
the likelihood that STSs will be generated experimentally.

The co-propagation of the FF and SH waves in SHG media is
described by known equations \cite{Kanashov,Malomed97},
\begin{eqnarray}
iu_{\xi }+\nabla _{\bot }^{2}u+u_{\tau \tau }-u+u^{\ast }v &=&0,  \label{FF}
\\
2iv_{\xi }+\nabla _{\bot }^{2}v+\delta v_{\tau \tau }-i\sigma v_{\tau
}-\gamma v+\left( 1/2\right) u^{2} &=&0.  \label{SH}
\end{eqnarray}
Here, $u=\sqrt{2}\mathcal{E}_{1}e^{-iz/z_{0}}z_{0}\omega
_{1}^{2}\chi ^{(2)}/(k_{1}c^{2})$ and
$v=\mathcal{E}_{2}e^{-2iz/z_{0}}z_{0}\omega _{1}^{2}\chi
^{(2)}/(k_{1}c^{2})$, $\mathcal{E}_{1,2}$ are complex
electric-field envelopes at the frequencies $\omega _{1}$ and
$2\omega _{1}$, $k_{1}\equiv k\left( \omega _{1}\right)$ is the FF
carrier wavenumber, $\xi =z/z_{0}$ and $\tau =(t-z/v_{g})/t_{0}$
are the normalized propagation distance and reduced time, $v_{g}$
and $t_{0}$ being the group velocity and time scale at FF,
$z_{0}=2t_{0}^{2}/k_{1}^{^{\prime \prime }}$, where
$k_{1}^{^{\prime \prime }}$ is the GVD coefficient at FF, and the
transverse radial coordinate is $r=\sqrt{2k_{1}\left(
x^{2}+y^{2}\right) /z_{0}}$. The diffraction operator, $\nabla
_{\bot }^{2}\equiv \partial ^{2}/\partial r^{2}+[(D-2)/r]\partial
/\partial r$, where $D=3$ or $2$ is the spatial dimension, acts on
the transverse coordinates $\{x,y\}\equiv \sqrt{
2k_{1}/z_{0}}\mathbf{r}$. In the experiment, units of time and
transverse and propagation distance in the scaled equations
typically correspond to $100 $ fs, $50$\thinspace $\mu $m, and $1$
cm, respectively \cite{Liu99}. Further, $\sigma $ in Eq.
(\ref{SH}) is GVM, and $\gamma \equiv 4+2z_{0}[2k(\omega
_{1})-k(2\omega _{1})]$ is the phase mismatch between the
harmonics. The equations conserve the Hamiltonian, momentum, axial
angular momentum, and the Manley-Rowe invariant (energy), which is
$Q=2\pi \int_{0}^{\infty }rdr\int_{-\infty }^{+\infty }d\tau
\,\left( |u|^{2}+4|v|^{2}\right) $ in the 3D case.

In Eq. (\ref{FF}), the FH dispersion is anomalous, while the ratio
$k_{2}^{^{\prime \prime }}/k_{1}^{^{\prime \prime }}\equiv \delta
$ of the SH and FH GVD coefficients is assumed negative,
corresponding to \emph{normal} GVD at SH. This case, which
corresponds to the experimental reality, has not yet been studied
(except for VA, with $\sigma =0$, in Ref. \cite{Malomed97}).

\begin{figure}[h]
\centerline{\includegraphics[width=75mm,angle=0,keepaspectratio]{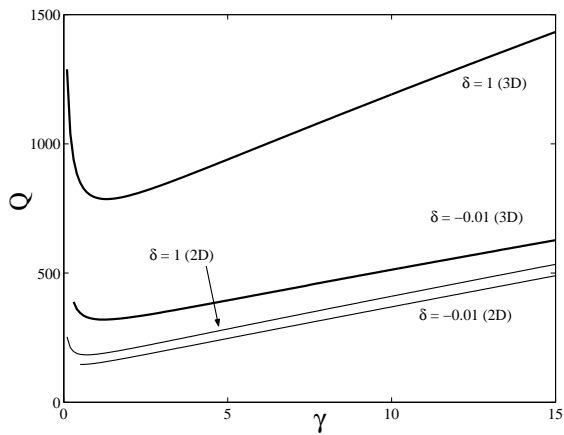}}
\caption{The energy of stationary spatiotemporal solitons vs. the
phase mismatch $\protect\gamma$ in the case of normal GVD at the
second harmonic ($\delta =-0.01$). The curves with
$\protect\delta=1$ represent earlier known solutions and are
included for comparison.} \label{qb}
\end{figure}

First, an estimate for the existence of STSs can be obtained for
small $|\delta |$. To this end, Eq. (\ref{SH}) is treated as a
linear equation for SH, driven by the term $\left( 1/2\right)
u^{2}$. A SH field component of the CW (continuous-wave) form,
$B_{\mathrm{cw}}\exp \left( i\mathbf{k\cdot r} -i\omega \tau
\right) $, which does not vanish as $r,|\tau |\rightarrow \infty $
and thus prevents the existence of the soliton, is determined by
zeros of the denominator of the corresponding Green's function,
which are located at
\begin{equation}
|\delta |\omega ^{2}-\sigma \omega -k^{2}-\gamma =0.  \label{omega}
\end{equation}
On the other hand, an asymptotic form of the solution to Eq.
(\ref{FF}) at $r,|\tau |\rightarrow \infty $ is $u\sim \rho
^{-(D-1)/2}\exp \left( -\rho \right) $, with $\rho \equiv
\sqrt{\tau ^{2}+r^{2}}$.

\begin{figure}[h]
\centerline{
\includegraphics[width=75mm,angle=0,keepaspectratio]{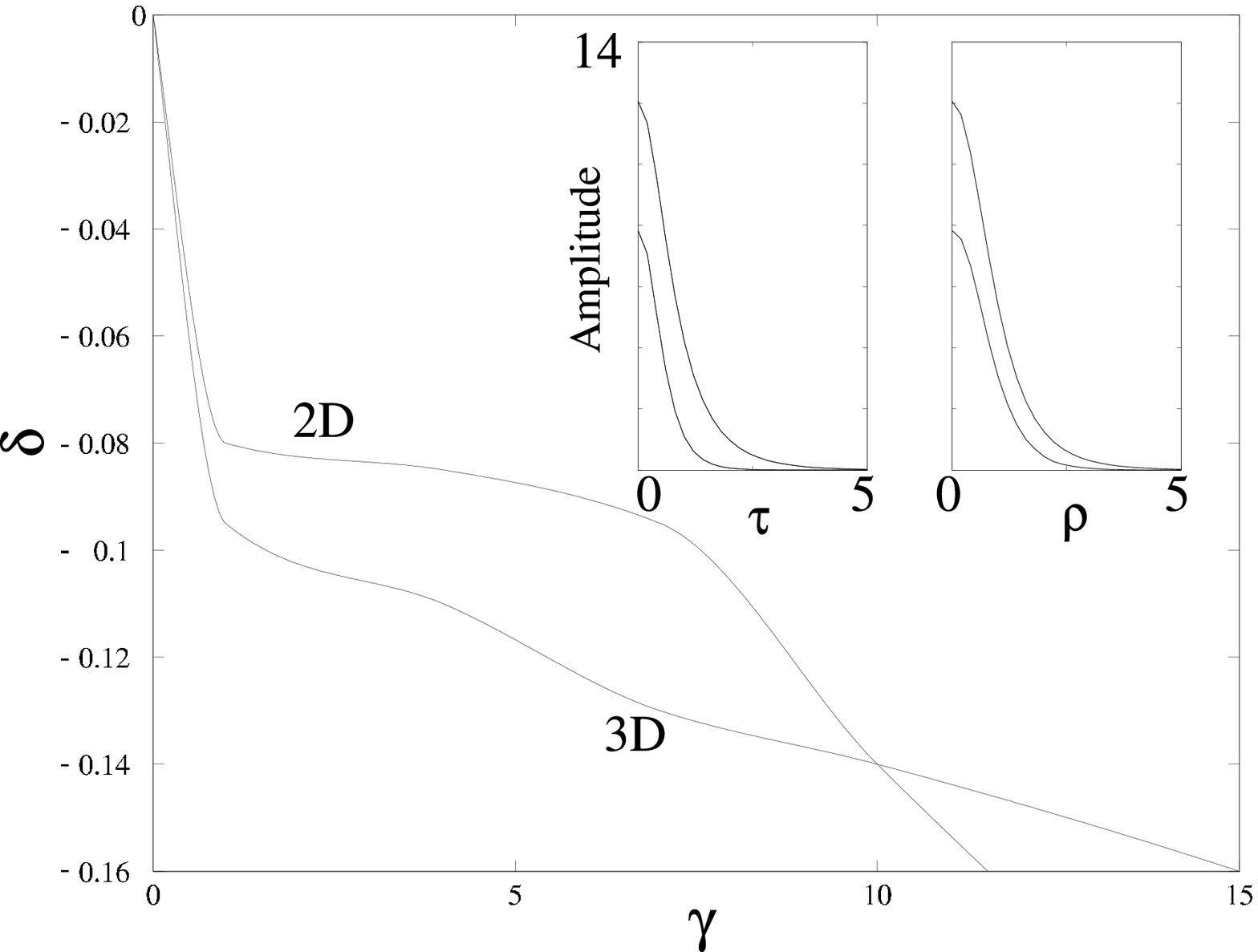}}
\caption{Stability regions for 2D and 3D solitons without walkoff
($\sigma = 0$). The solitons are stable above the respective
borders. The insets show central profiles of the 3D soliton in the
temporal and spatial directions, for $\delta = -0.02$ and
$\gamma=5$. The upper and lower curves pertain to the FF and SH
components $u$ and $v$, respectively.} \label{domain}
\end{figure}

An estimate for the maximum amplitude $B_{\mathrm{cw}}$ of the
possible CW component can be derived. To simplify the
presentation, one can adopt a model form of the FF field complying
with the above asymptotics and securing a globally correct shape
of the solution. In the 3D case, it is $u=\mathrm{ const\cdot
}\rho ^{-1}\left( \sinh \rho \right) \mathrm{sech}^{2}\rho $, but
the final exponential estimate (see below) does not predicate on
this ``ansatz''. Straightforward calculations [Fourier transform
of $u^{2} (\mathbf{r},\tau )$, inverse transform of its product
with the Green's function of Eq. (\ref{SH}), and isolating a CW
contribution from $\omega$ determined by Eq. (\ref{omega})] show
that, in the case $\sigma =0$ (no-walkoff), the CW intensity is
exponentially small (for small $|\delta |$) if $\gamma >0$, an
upper estimate for it being
\begin{equation}
B_{\mathrm{cw}}^{2}<\mathrm{const}\cdot \gamma \left( \gamma /|\delta
|\right) ^{3/2}\exp \left( -\pi \sqrt{\gamma /|\delta |}\right) .
\label{estimate}
\end{equation}
While the pre-exponential part of this estimate may be
ansatz-dependent, the crucially important exponential part is not,
and it applies to the 2D case as well; if $\sigma \neq 0$, it is
replaced by $\exp \left\{ -\pi \left[ \sqrt{\left( \gamma /|\delta
|\right) +\left( \sigma /2\delta \right) ^{2}} -\left| \sigma
/2\delta \right| \right] \right\} $, i.e., the walkoff makes the
estimate weaker.

Thus, although it is not guaranteed that the STS exists as a
rigorous solitary-wave solution, the one with the CW tail obeying
the exponentially small estimate may be tantamount to a true
soliton in any possible experiment. The actual region of the
existence and stability of STSs should be found numerically, which
is done below.

\begin{figure}[h]
\centerline{
\includegraphics[width=75mm,angle=0,keepaspectratio]{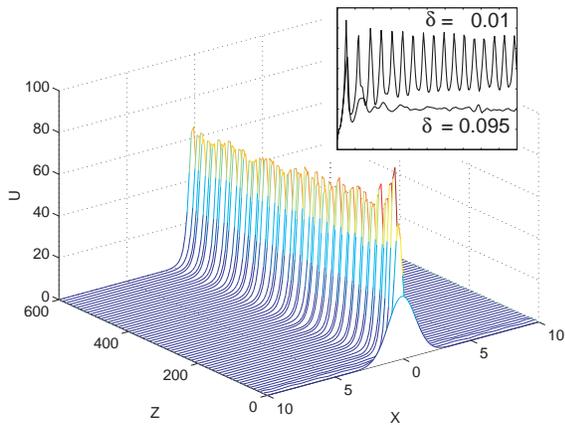}}
\caption{Evolution of the 3D soliton self-trapping from an initial
Gaussian pulse for $\delta =-0.01,\gamma=1$, and $\sigma=0$. The
propagation distance corresponds to $\approx 800$ dispersion
lengths. The inset shows the evolution of the FF amplitude on a
shorter scale, for the same case and the one with $\delta =0.095$,
closer to the instability border. } \label{propagation3D}
\end{figure}

Stationary solutions were found by solving the $z$-independent
version of Eqs. (\ref{FF}) and (\ref{SH}) with $\delta <0$.
Although the above analysis suggests that STS may have a small
nonvanishing tail in the SH component, no tails were found, with
the numerical accuracy available [$\mathcal{O}(10^{-4})$], in all the pulses 
which were identified as solitons. It was also verified that the shape of the
soliton does not change with increasing the numerical accuracy
and/or size of the integration domain (for instance, there was no
change with the increase of the number of points from $64^{3}$ to
$256^{3}$). Insets in Fig. \ref{domain} show an example of a 3D
soliton, found for $\delta =-0.02$, $\gamma =5$ and $\sigma =0$.
An STS family was generated by varying $\delta$ and $\gamma$, Fig.
\ref{qb} showing the soliton's energy versus $\gamma $. A
conclusion of direct relevance to experiment is that, in the most
challenging 3D case, energy required to create a soliton with
normal GVD at SH is \emph{smaller} than in the previously studied
case $\delta =1$ by a factor $\simeq 2.5$.

Further consideration shows that, at large $\gamma $, the bulk of
the STS's energy is in FF (in compliance with the cascading limit
\cite{Buryak}), while at smaller $\gamma $ the amplitudes of both
components are nearly equal, hence SH has roughly four times the
energy of the fundamental. This shows that, remarkably, the
soliton is able to exist keeping $\simeq 80\%$ of its energy in
the \emph{normal}-GVD component.

A crucially important issue is stability of the solitons. First of
all, it may be judged using the Vakhitov-Kolokolov criterion,
which states that solitons may only be stable if $dQ/d\gamma >0$
\cite{Buryak}. Figure \ref{qb} shows that the soliton families
satisfy this criterion, unless $\gamma $ is too small. A
systematic stability test was performed by direct simulations of
Eqs. (\ref{FF}) and (\ref{SH}). The results are summarized in Fig.
\ref {domain}.

At small $|\delta|$, the stability border in Fig. \ref{domain} is
linear, which complies with the fact that the exponential factor
in the estimate (\ref{estimate}) is a function of the ratio
$\gamma/|\delta|$ (the exponential estimate yields a factor $\sim
10^{-5}$ at the stability border in the figure). Furthermore, the
2D and 3D stability borders are almost identical at small $|\delta
|$, in agreement with the fact that the exponential estimate is
the same for both dimensions. The solitons are truly robust: not
only they are not destroyed by perturbations, but they also
readily self-trap from initial pulses of quite an arbitrary shape,
which is illustrated by Fig. \ref {propagation3D}. Note that the
transition from the initial profile to the soliton's one takes the
propagation distance of few dispersion lengths ($z_{D}$), and the
STS remains stable over an extremely long distance $\approx
800z_{D}$. This shows not only that the STS will be stable in any
experimental setup, but also that possible energy leakage due to
the formation of the above-mentioned CW tail cannot be spotted in
the course of the extremely long propagation. The same figure
shows persistent intrinsic oscillations in the stable STS, which
is attributed to excitation of an intrinsic mode. The excitation
is tangible if the initial pulse is launched at a point which is
located deep within the stability region. If the pulse is taken
close to the stability border, the inset to Fig.
\ref{propagation3D} shows that the vibrations quickly fade, the
pulse relaxing to a stationary soliton through transient emission
of radiation (the same is observed in the 2D case).

\begin{figure}[h]
\centerline{\includegraphics[width=75mm,angle=0,keepaspectratio]{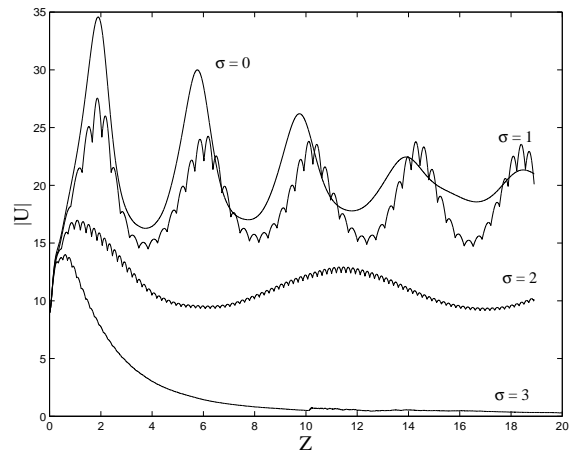}}
\caption{The evolution of the FF amplitude for 3D solitons with
$\protect\gamma =15,\protect\delta =-0.1$ and various values of
the GVD parameter $\protect\sigma$. The soliton is generated from
a Gaussian pulse launched in the FF component. It exists up to
$\protect\sigma =3$.} \label{walkoff2}
\end{figure}

The realistic model must include GVM. A previous work
\cite{Mihalache99a} addressed ``walking'' STSs, but in the case
$\delta >0$. Starting with a Gaussian pulse in the FF component,
we tried to generate STSs with a finite GVD parameter $\sigma $.
Figure \ref{walkoff2} shows the result: at large $\gamma $ and
relatively large negative $\delta $, the bullet self-traps and is
found to be fully localized for moderate values of $\sigma $. The
pulse cannot self-trap (decaying into radiation) if $\sigma >2$.
The stability diagram displayed in Fig. \ref{domain} can be
extended, adding $\sigma $ as a parameter. For $\sigma \,\,_{\sim
}^{<}\,\,1$, the stability region is nearly the same as at $\sigma
=0$, but with the increase of $\sigma $, it quickly shrinks and no
longer exists for $\sigma >2$.

Lastly, stationary 3D solitons with an internal vorticity were
found too. However, they are always unstable against perturbations
breaking their axial symmetry, similar to what was found for the
same type of STS in the case $\delta >0$ \cite{Mihalache2000}.

In conclusion, we have shown in the numerical form that 2D and 3D
stable spatiotemporal solitons exist in quadratically nonlinear
media when the dispersion is anomalous at FF but normal at SH. An
exponential analytical estimate for the existence of the solitons
was obtained. The solitons readily self-trap from a Gaussian pulse
launched at FF, showing no trace of decay over extremely long
propagation distance. The solitons have the energy smaller than
their counterparts in the case of the anomalous dispersion at SH,
and they exhibit robustness to the walkoff between the harmonics.
These features imply that the solitons of this type can be created
under a variety of experimental conditions.

We appreciate a valuable discussion with D. Mihalache and K.
Beckwitt. The work was supported in a part by the Binational
(US-Israel) Science Foundation (contract No. 1999459) and by a
matching grant from the Tel Aviv University (TAU). Access to
supercomputing facilities was provided by the High Performance
Computing Unit at TAU.


\begin{thebibliography}{99}
\bibitem{Yaron}  Y. Silberberg, \ol\textbf{15}, 1281 (1990).

\bibitem{review}  G.I. Stegeman and M. Segev, Science \textbf{286}, 1518
(1999).

\bibitem{1d}  R. Schiek, Y. Baek, and G.I. Stegeman, \pre {\bf 53}, 1138
(1996).

\bibitem{2d}  W.E. Torruellas \textit{et al.}, \prl {\bf 74}, 5036 (1995).

\bibitem{Liu99}  X. Liu, L.J. Qian, and F.W. Wise, \prl {\bf 82}, 4631
(1999); X. Liu, K. Beckwitt, and F.W. Wise, \pre {\bf 61}, R4722 (2000); X.
Liu, K. Beckwitt, and F.W. Wise, \prl {\bf 85}, 1871 (2000).

\bibitem{Macleod}  R. McLeod, K. Wagner, and S. Blair, \pra {\bf 52}, 3254
(1995).

\bibitem{Rasmussen}  J.J. Rasmussen and K. Rypdal, Phys. Scripta
\textbf{33}, 481 (1986).

\bibitem{Blagoeva}  A.B. Blagoeva \textit{et al.}, \jqe {\bf QE-27}, 2060
(1991).

\bibitem{Eisenberg}  H.S. Eisenberg \textit{et al.}, \prl {\bf 87}, 043902
(2001).

\bibitem{Buryak}  C. Etrich \textit{et al}., Progr. Opt. \textbf{41}, 483
(2000); A.V. Buryak \textit{et al.}, Phys. Rep. \textbf{370}, 63 (2002).

\bibitem{Kanashov}  A.A. Kanashov and A.M. Rubenchik, Physica D \textbf{4},
122 (1981).

\bibitem{Skr}  D.V. Skryabin and W.J. Firth, \oc\textbf{148}, 79 (1998).

\bibitem{Malomed97}  B.A. Malomed \textit{et al.}, \pre {\bf 56}, 4725
(1997).

\bibitem{Mihalache98}  D. Mihalache \textit{et al.}, \oc {\bf 152}, 365
(1998); D. Mihalache \textit{et al.}, \oc {\bf 159}, 129 (1999).

\bibitem{Mihalache99a}  D. Mihalache \textit{et al.}, \oc {\bf 169}, 341
(1999); D. Mihalache \textit{et al.}, \pre {\bf 62}, 7340 (2000).

\bibitem{Mihalache2000}  D. Mihalache \textit{et al.}, \pre\textbf{\ 62},
R1505 (2000).
\end{thebibliography}
\end{document}